# Enhanced optical nonlinearities in air-cladding silicon pedestal waveguides

YAOJING ZHANG,[1] ZHENZHOU CHENG,[2] YIFEI YAO,[3] HON KI TSANG[1,*]

[1] Department of Electronic Engineering, The Chinese University of Hong Kong, Shatin, New Territories, Hong Kong, People's Republic of China
[2] Department of Chemistry, The University of Tokyo, Tokyo 113-0033, Japan
[3] Division of Biomedical Engineering, The Chinese University of Hong Kong, Shatin, New Territories, Hong Kong, People's Republic of China
*Corresponding author: hktsang@ee.cuhk.edu.hk

**The third-order optical nonlinearity in optical waveguides has found applications in optical switching, optical wavelength conversion, optical frequency comb generation, and ultrafast optical signal processing. The development of an integrated waveguide platform with a high nonlinearity is therefore important for nonlinear integrated photonics. Here, we report the observation of an enhancement in the nonlinearity of an air-cladding silicon pedestal waveguide.  We observe enhanced nonlinear spectral broadening compared to a conventional silicon-on-insulator waveguide. At the center wavelength of 1555 nm, the nonlinear-index coefficient of air-cladding silicon pedestal waveguide is measured to be about 5% larger than that of a conventional silicon-on-insulator waveguide. We observe enhanced spectral broadening from self-phase modulation of an optical pulse in the pedestal waveguide. The interaction of light with the confined acoustic phonons in the pedestal structure gives rise to a larger nonlinear-index coefficient. The experimental results agree well with the theoretical models.**

Silicon waveguides fabricated on silicon-on-insulator (SOI) wafers have been extensively studied for nonlinear integrated photonics [1-3]. As a nonlinear material, the top layer single-crystal silicon exhibits a third-order optical nonlinear susceptibility which is more than two orders of magnitude larger than that in optical fibers in the telecommunication band [4]. The large refractive index difference between the top layer silicon and the buried-oxide (BOX) substrate can further enhance the effective optical nonlinearity because of the tight optical confinement in the vertical direction, which allows the effective area of silicon waveguide to be another two orders of magnitude smaller than optical fibers.  Many applications of third-order optical nonlinearity have been widely studied in SOI devices, such as optical parametric amplification [5], all-optical wavelength conversion [6] and supercontinuum generation [7].

It is possible to increase the nonlinear conversion efficiency and bandwidth of nonlinear silicon devices by optical dispersion engineering [8], reducing the nonlinear loss by removing the free carriers produced by two photon absorption [9], and the use of high-quality-factor optical resonators to enhance the intensity of light [10]. In recent years, new materials and structures have also attracted increasing attention as a means to attain higher nonlinearity than the conventional SOI waveguide. Examples include the integration of graphene on top of silicon to increase the four wave mixing conversion efficiency [11], the use of amorphous silicon waveguides for higher third-order nonlinear coefficient than crystalline silicon waveguide [12] and the use of molybdenum disulphide layers over the silicon waveguide to enhance the Kerr nonlinearity [13]. The studies on new integrated structures and materials are of significance in further advancing the use of silicon waveguide based structures for nonlinear photonics.

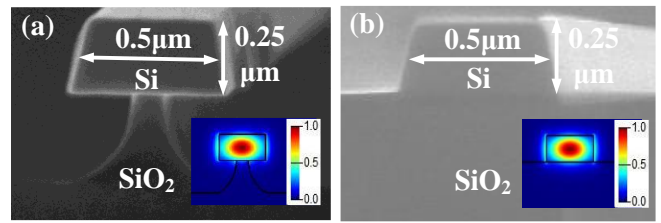

Fig. 1. Scanning electron microscope (SEM) images of fabricated structures. (a) Silicon pedestal waveguide. (b) SOI waveguide. Inset: correspondingly electric field distributions of transverse-electric (TE) mode in waveguides.

In this paper we consider the possible optomechanical enhancement of optical nonlinearity in silicon nanoscale structures.  Electrostriction, which is the variation of material density under an applied electric field, can produce a change in refractive index at timescales similar to the transit time for the acoustic wave to move across the optical mode [14]. Typically, in optical fibers, the optical nonlinearity from electrostriction has a much slower response time than the optical Kerr effect, and thus different effective nonlinear-index coefficient are observed for picosecond pulses and microsecond pulses [14]. In optical fibers, it has been demonstrated that the electrostriction could contribute to significantly higher effective nonlinear-index coefficient [14]. However, there has not been any experimental investigation of possible enhancement from electrostriction in silicon waveguides.

In this Letter, we experimentally study the third-order optical nonlinearity of an air-cladding silicon pedestal waveguide (SPW). The buried oxide (BOX) layer underneath the top silicon waveguide is partially removed by using hydrofluoric acid (HF) solution, as shown in Fig 1. (a). By using the nonlinear spectral broadening of short optical pulses, the nonlinear-index coefficient is measured to be $(3.15 \pm 0.3) \times 10^{-18}$ m$^2$/W in the SPW, which is about 5% larger than that of a conventional SOI waveguide. The result can be possibly explained by the elastic wave induced nonlinear-index coefficient enhancement. We apply two methods to calculate the enhancement and both agree well with the experimental results.

The samples are fabricated on a commercial SOI wafer (SOITEC Inc.) which has a 0.25 μm top silicon layer and a 3 μm BOX underneath the top silicon layer. We fabricate two types of channel waveguides by dry etching to the buried oxide: a SPW and a SOI waveguide, as shown in Fig. 1. The widths and lengths of all waveguides are fixed as 0.5 μm and 484 μm, respectively. We use the 10% HF solution to remove the BOX below the top-layer silicon devices in the SPW. The BOX in SPW device is etched in the HF solution for 330 seconds. The BOX below the silicon in the SOI waveguide is not removed, and the SOI waveguide is used as a control sample for comparison in the experimental measurement. The SPW and SOI waveguide are designed as single mode and have the same physical dimensions as shown in Fig.1. The calculated electrical field distributions of the TE mode are also shown in Fig. 1.

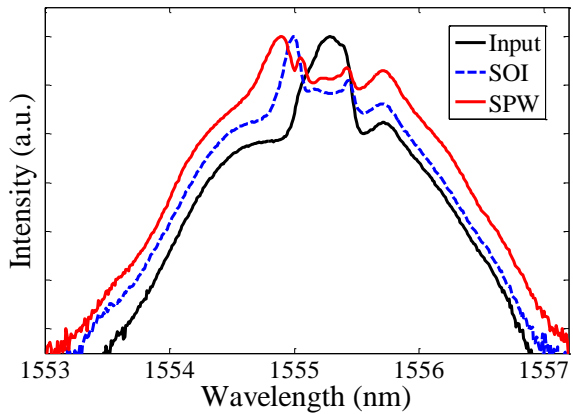

Fig. 2. Comparison of measured spectra of input pulse and transmission pulses from SOI waveguide and SPW.

We first experimentally characterize the spectral broadening in the two devices. A chirped pulse is produced from a gain-switch distributed-feedback (DFB) laser which is driven by an electrical impulse generator at a repetition rate of 1 MHz. The optical pulses have a full width half-maximum pulse width of about 60 ps and a center wavelength of 1555 nm, and are used as the input for the spectral broadening measurements [15]. The pulses are coupled into and out of the waveguides via subwavelength waveguide gratings [16] which have 3 dB optical bandwidth of 40 nm. Fiber to waveguide to fiber insertion losses are measured as 18 dB at wavelength of 1550 nm. By comparing the input pulse spectrum and the transmitted pulse spectra of two devices, we obtain a larger spectral broadening in the SPW as indicated by the red solid line in Fig.2.

As shown in Fig. 2, nonlinear spectral broadening from self-phase modulation is observed in the output spectrum of the SPW [17]. The new frequency components are blue-chirped near the rising edge and red-chirped near the falling edge of the pulses as they propagate through the waveguide The measured spectral broadening appears wider at the short wavelength side produced by the rising edge of the pulse because the waveguide has excess nonlinear losses from free carriers generated by two photon absorption. The free carrier absorption at the trailing edge of the pulse reduces the output power of the longer wavelength spectral broadening. On the other hand, the positive frequency chirp (blue shift) generated by the rising edge of the pulse does not experience free carrier absorption as the free carrier population has not yet been produced by the pulse at the rising edge.

Since the interaction of self-phase modulation with optical dispersion can have an influence on the magnitude of spectral broadening, to clarify the mechanism, we theoretically simulate the dispersion in the SPW and the SOI waveguide by using a commercial software. According to the simulation, the waveguide dispersion in SPW is 2600 ps/nm/km and the group-velocity dispersion (GVD) coefficient $\beta_2$ is calculated to be -3 × 10$^{-24}$, using the result $D = -2\pi \cdot c \cdot \beta_2 / \lambda^2$ [15], where c is the velocity of light in vacuum and λ is the wavelength in vacuum. The dispersion length is given by $L_D = T_0^2 / |\beta_2|$ [15], and is thus calculated to be several kilometers, and therefore much greater than the waveguide length. A similar dispersion length is obtained for the SOI waveguide. Therefore, for the devices in our study, dispersion does not play a significant role in the measured spectral broadening of the waveguides.

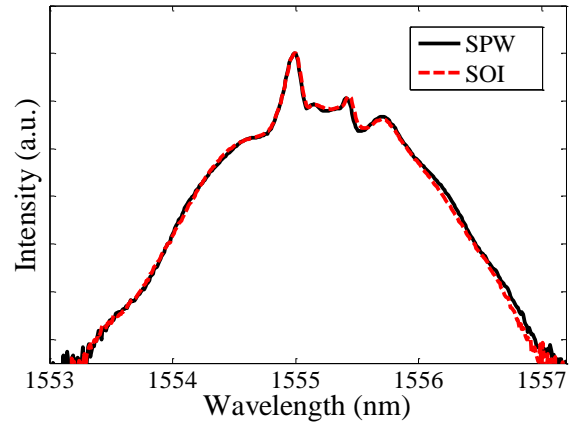

Fig. 3. Fitting spectra between the SPW and the SOI waveguide. Both waveguides use the same grating design to ensure the equal coupling efficiency and spectral bandwidth.

We previously used the single-mode fiber (SMF) with the known nonlinear-index coefficient and effective area to match the spectra from SOI waveguides to measure the nonlinear-index coefficient of the SOI waveguides [4]. In this method, different lengths of SMFs are carefully tailored for fitting the same transmission spectra. In this paper we determine the nonlinear-index coefficient of SPW by comparing with the spectral broadening obtained from the SOI waveguide. The experimental comparison involved varying the input power to the SPW until measured spectral broadening is identical to that obtained for the SOI waveguide at a different input power. Both the SOI and SPW measurements employ the same waveguide grating coupler design, and this ensured that they have same coupling efficiency and spectral characteristics.

After observing the broader output spectrum from the SPW with the same input power as the SOI waveguide is used (Fig. 2), we decrease the input power of the SPW until the measured output spectral broadening matches exactly with the output spectrum of the SOI waveguide at the original power level. When we decrease the power to about 0.3W of peak-coupled power of the SPW, we observe well matched spectra between the SPW and the SOI waveguide as shown in Fig. 3. This means that two waveguides have the same phase shifts Δφ. We calculate the effective areas of the SPW and SOI waveguide by using FDTD solutions (Lumerical Inc.) software as $6.76 \times 10^{-14}$ m$^2$ and $6.80 \times 10^{-14}$ m$^2$. The phase shift is given by $\Delta\varphi = 2\pi \cdot n_2 \cdot P \cdot L_{eff} / \lambda / A_{eff}$, where $n_2$ is the nonlinear-index coefficient, P is the peak power, $L_{eff}$ is the effective length of waveguide, λ is the wavelength in vacuum and $A_{eff}$ is the waveguide effective area [15]. Using the nonlinear-index coefficient of the SOI as $3 \times 10^{-18}$ m$^2$/W [18], we thus obtain nonlinear-index coefficient of the SPW to be $(3.15 \pm 0.3) \times 10^{-18}$ m$^2$/W which includes an estimated 10% uncertainty in the measurement of absolute peak power. The relative powers coupled into the SOI and SPW as measured from the incident and output powers, are accurate to about 2%.

To explain the possible mechanism for the observed enhancement in the SPW, we first note that the acoustic phonons from Brillouin scattering can cross the optical mode in about 59 picoseconds [3], which is comparable to the pulse width used. We can apply the Kramers-Kronig relation to calculate the change in refractive index from stimulated Brillouin scattering (SBS) loss and gain as follows,

$$\Delta n(\Omega) = \frac{c}{\pi} \mathcal{P} \int_0^\infty \frac{\Delta\alpha(\Omega)d\Omega}{\Omega^2 - \omega^2}. \qquad (1)$$

In equation (1), c is the speed of light in vacuum, $\mathcal{P}$ denotes taking the Cauchy principal value, Ω and ω are the angular frequencies, and Δα is the change in the absorption coefficient [19]. In equation (1), Δα is negative for gain at the Stokes-shifted wavelength and positive for the loss at the anti-Stokes shifted wavelength. By using the Lorentzian Brillouin gain function [3] we can estimate Δα using

$$\Delta\alpha(\Omega) = PG(\Omega) = P \frac{2\gamma_{SBS}}{4(\frac{\Omega - \Omega_m}{\Gamma_m})^2 + 1}, \qquad (2)$$

where P is the optical power, $\Gamma_m/2\pi$ = 30 MHz is the acoustic linewidth, $\Omega_m/2\pi$ = 8.2 GHz is the frequency shift and the Brillouin gain coefficient $2\gamma_{SBS}$ = 3218 W$^{-1}$m$^{-1}$ for the SPWs [3]. The contribution to the nonlinear-index coefficient from the optomechanical interaction with the elastic wave $n_{2e}$ can be obtained from equation (1) and using

$$\Delta n = n_{2e}I = n_{2e}\frac{P}{A_{eff}}, \qquad (3)$$

$n_{2e}$ is calculated as $0.15 \times 10^{-18}$ m$^2$/W (with $A_{eff}$ as $6.76 \times 10^{-14}$ m$^2$, I as the intensity of light) and the total nonlinear-index coefficient is thus $3.15 \times 10^{-18}$ m$^2$/W in agreement with the experiment.

We also considered another model to estimate the nonlinear-index coefficient resulting from the acoustic waves. For nanoscale waveguides, it has been demonstrated that both electrostriction and radiation pressure effects will generate optical forces [20]. Electrostriction arises from material photoelastic tensor, while radiation pressure is from discontinuous dielectric boundaries of the waveguide. It has been demonstrated that the radiation pressure can generate a large force in nanoscale waveguide for its large field enhancement and high confinement [20]. We estimate the nonlinear-index coefficient produced by electrostriction and radiation pressure as follows. The power normalized total stress $\sigma_{kl}^{opt}$ consisting of both electrostriction and radiation pressure may be estimated to be $1.2 \times 10^{-8}$ N/W and $0.2 \times 10^{-8}$ N/W in lateral and vertical boundaries of cross section of the waveguide [20].

We can obtain the strain-induced change in the inverse dielectric tensor $\Delta(\varepsilon_{ij}^{-1})$ as follows,

$$\Delta(\varepsilon_{ij}^{-1}) = p_{ijkl}S_{kl}, \qquad (4)$$

where the corresponding strain $S_{kl}$ is calculated as follows,

$$S_{kl} = C_{klmn}\sigma_{mn}^{opt}, \qquad (5)$$

where $C_{klmn}$ is the elastic compliance tensor [20].

Then, the corresponding refractive index change $\Delta n$ and nonlinear-index coefficient $n_{2e}$ can be calculated as follows,

$$\Delta n = -\frac{n^3}{2}\Delta(\varepsilon^{-1}), \qquad (6)$$

$$|n_{2e}| = \left|\frac{2\Delta n \eta_0}{n|E|^2}\right|, \qquad (7)$$

where $\eta_0$ is the free-space impedance as 377 Ω [14]. With equation (6), the elastic waves contributed to the nonlinear-index coefficient is calculated as $0.15 \times 10^{-18}$ m$^2$/W which is in agreement with the Kramers-Kronig calculation and the experimental results. In the case of the SOI waveguide, the effect of elastic waves on the nonlinear-index coefficient is absent because of the large leakage of acoustic phonons to the silica substrate resulting in short phonon lifetime in the SOI waveguide [3].

In conclusion, we study the spectral broadening of an air-cladding SPW and observe enhanced self-phase modulation which could not be accounted for just by the difference in effective areas of the SOI waveguide and SPW. The SPW has an enhanced nonlinear-index coefficient for the 60 picosecond pulses because of the interaction with acoustic phonons and elastic waves. The measurement results agree with theoretical estimates of the enhanced nonlinear-index coefficient from SBS and also with calculations from considering the material photoelastic tensor and different compliance tensors for the pedestal waveguide.

**Funding.** Hong Kong Research Grants Council (RGC) (416913).

**Acknowledgment**. We thank Zejie Yu for helping with simulation, Jie Liu and Qijie Xie for helping with use of experimental instruments.